\documentclass[]{aa}  

\usepackage{graphicx}
\usepackage{inputenc}
\usepackage{txfonts}
\usepackage{natbib}
\bibpunct{(}{)}{;}{a}{}{,}

\begin{document} 

   \title{A radio-map of the colliding winds in the very massive binary system HD\,93129A}

   \author{P. Benaglia
          \inst{1}\fnmsep\thanks{Fellow of CONICET},
      B. Marcote\inst{2},
     J. Mold\' on\inst{3},
     E. Nelan\inst{4},
     M. De Becker\inst{5},
     S. M. Dougherty\inst{6},	
          \and
          B. S. Koribalski\inst{7}
          }

   \institute{Instituto Argentino de Radioastronom\'{\i}a, CCT-La Plata, CONICET,
              Argentina; \email{paula@iar-conicet.gov.ar}
       \and Departament d'Astronomia i Meteorologia, Institut de Ci\`encies del
Cosmos, Universitat de Barcelona, IEEC-UB, Mart\'{\i} i Franqu\`es 1,
E08028 Barcelona, Spain
      \and ASTRON Netherlands Institute for Radio Astronomy, Postbus 2, 7990AA Dwingeloo, The Netherlands
      \and Space Telescope Science Institute, USA
     \and  Department of Astrophysics, Geophysics and Oceanography, University of
Li\`ege, 17 All\'ee du 6 Ao\^ut B5c, B-4000 Sart Tilman, Belgium
     \and NRC Herzberg Astronomy and Astrophysics, DRAO, PO Box 248, Penticton B.C., Canada V0H 1K0
    \and Australia Telescope National Facility, CSIRO Astronomy \& Space Science
   P O Box 76, Epping, NSW 1710, Australia
             }

 \date{Received ; accepted }

 \abstract {Radio observations are an effective tool to discover
   particle acceleration regions in colliding-wind binaries, through
   detection of synchrotron radiation; these regions are natural
   laboratories for the study of relativistic
   particles. Wind-collision region (WCR) models can reproduce the
   radio continuum spectra of massive binaries that contain both
   thermal and non-thermal radio emission; however, key constraints
   for models come from high-resolution imaging. Only five WCRs have
   been resolved to date at radio frequencies at milliarcsec (mas)
   angular scales. The source HD\,93129A, prototype of the very few
   known O2 I stars, is a promising target for study: recently, a
   second massive, early-type star about 50 mas away was discovered,
   and a non-thermal radio source detected in the region. Preliminary
   long-baseline array data suggest that a significant fraction of
   the radio emission from the system comes from a putative WCR.}
   {We sought evidence that HD\,93129A is a massive binary system
   with colliding stellar winds that produce non-thermal radiation,
   through spatially resolved images of the radio emitting regions.}
 {We completed observations with the
   Australian Long Baseline Array (LBA) to resolve the system at mas
   angular resolutions and reduced archival Australia Telescope
   Compact Array (ATCA) data to derive the total radio emission. We also
   compiled optical astrometric data of the system in a homogeneous
   way. We reduced historical Hubble Space Telescope data and
   obtained absolute and relative astrometry with milliarcsec
   accuracy.}

   {The astrometric analysis leads us to conclude that the two
   stars in HD\,93129A form a gravitationally bound system. The LBA
   data reveal an extended arc-shaped non-thermal source between the
   two stars, indicative of a WCR. The wind momentum-rate
   ratio of the two stellar winds is estimated.  The ATCA data
   show a point source with a change in flux level between 2003-4
   and 2008-9, that is modeled with a non-thermal power-law
   spectrum with spectral indices of $-1.03 \pm 0.09$ and $-1.21
   \pm 0.03$ respectively.  The mass-loss rates derived from the
   deduced thermal radio emission and from the characteristics of
   the WCR are consistent with estimates derived by other authors.}
   {}

   \keywords{Stars: massive -- binaries: general -- Radio continuum:
   stars -- Radiation mechanisms: non-thermal -- stars: individual:
   HD\,93129A}

\titlerunning{The WCR of HD\,93129A}
\authorrunning{P. Benaglia et al.}

   \maketitle 


\section{Introduction} 

Wind-collision regions (WCR) in binary systems with two hot
massive stars constitute natural laboratories to study relativistic
particles \citep{EU1993,BR2003,PD2006,Debecker2007}, as do supernova
remnants but with somewhat different physical and geometrical
parameters. A WCR is the site of particle acceleration that generates
a population of relativistic electrons, as now demonstrated in more
than 40 massive binary systems. The particle-accelerating
colliding-wind binary (PACWB) status of these objects is in most cases
revealed by radio observations through the detection of synchrotron
emission \citep{pacwbcata}.

WCR models are able to reproduce the radio continuum spectra of these
massive binaries \citep{PD2006}. The radiation from these systems is
the superposition of  thermal emission from the stellar winds of
each binary component and the non-thermal emission from the WCR. A
key ingredient for constraining models of the non-thermal component is
expected to come from high-resolution radio imaging. The non-thermal
emission region associated with colliding winds has been mapped at
radio frequencies for WR\,140 \citep{Dougherty2005}, WR\,146
\citep{WCRwr146}, WR\,147 \citep{WCRwr147}, Cyg\,OB2\,\#5
\citep{Contreras1997,Ortiz2011} and Cyg\,OB2\,\#9 \citep{VLBIcyg9}. In
each case, the WCR is resolved by observations with angular resolutions
of a few milliarcseconds (mas).

HD\,93129A is the brightest source in the most crowded part of Tr 14 in the
Carina nebula and has been studied for several decades. The record 
of its special
spectral characteristics can be traced back to \citet{Payne1927} (see
also Table~\ref{datos-hd93} for information on other studies). It now 
is recognised as hosting the prototype O2\,If$+$ \citep{Walborn2002}. 
Radio emission from HD\,93129A has been detected between 1.4
and 24.5 GHz \citep[e.g.][]{Benaglia2006}, showing a flux decreasing
with frequency indicative of synchrotron radiation, most probably
produced in a WCR between stars in a colliding-wind binary system. The
Australian Long Baseline Array (LBA)\footnote {The Australian Long
Baseline Array is part of the Australia Telescope which is funded by
the Commonwealth of Australia for operation as a National Facility
managed by CSIRO.} observation presented by \citet{Benaglia2010}
suggested that a significant fraction of the radio emission comes from
a region compatible with a WCR.

In this paper, we offer an in-depth discussion of a more recent
LBA observation of HD\,93129A, aiming to provide compelling evidence
of the WCR origin for the bulk of the radio emission, in agreement
with the so-called standard scenario for PACWBs. These results allow
us to spatially resolve for the first time the synchrotron emission
region and to derive specific information about the non-thermal radio
contribution. The two component nature of HD\,93129A is presented,
including the most recent astrometric measurements that show the
relative position of the components over time (Sect.\,\ref{target}),
and a new analysis of Hubble Space Telescope/Fine Guidance Sensor
(HST/FGS) data that allow us to correlate the optical positions of
the two components with the position of the radio emission
(Sect.\,\ref{HST}). The LBA observations are presented in
Sect.\,\ref{lba}. The main results and general discussion are in
Sections \ref{results} and \ref{discussion}, with a final statement of
our conclusions in Sect.\,\ref{conclusions}.


\section{The HD\,93129A stellar system}\label{target}

The object HD\,93129A was first classified as an O3 star
\citep{Walborn1982}, though later was reclassified as the prototype O2
I star \citep{Walborn2002}. \citet{Mason1998} had suggested it was a
speckle binary, and was confirmed as having two components, Aa and Ab, 
from HST/FGS
observations \citep{Nelan2004}; these authors provide the separation
at the epoch of the observations, the difference in apparent visual
magnitude between the primary and the secondary and the position angle
($PA$) (see also \citealt{Nelan2010}). \citet{Maiz2008} detected
proper motion along the radius vector between the components
after analysis of data from 1996, 2002 and 2004: the binary
nature is supported by the relative velocity of about 2~mas~yr$^{-1}$, 
higher than the expected relative velocities in the field. The authors
suggested a highly elliptical or inclined orbit, or both, although
they did not determine the orbital parameters.  The companion is
assumed coeval with the primary and of the same spectral type as 
various other OB stars nearby (O3.5 V) (see Table~\ref{datos-hd93}).
Through detailed ultra-violet and H$\alpha$ line modeling, \cite{Taresch1997}
and \cite{Repolust2004} derived the main parameters of HD\,93129A (see
Table~\ref{datos-hd93}).

\begin{figure}
   \centering
\includegraphics[width=9cm,angle=0]{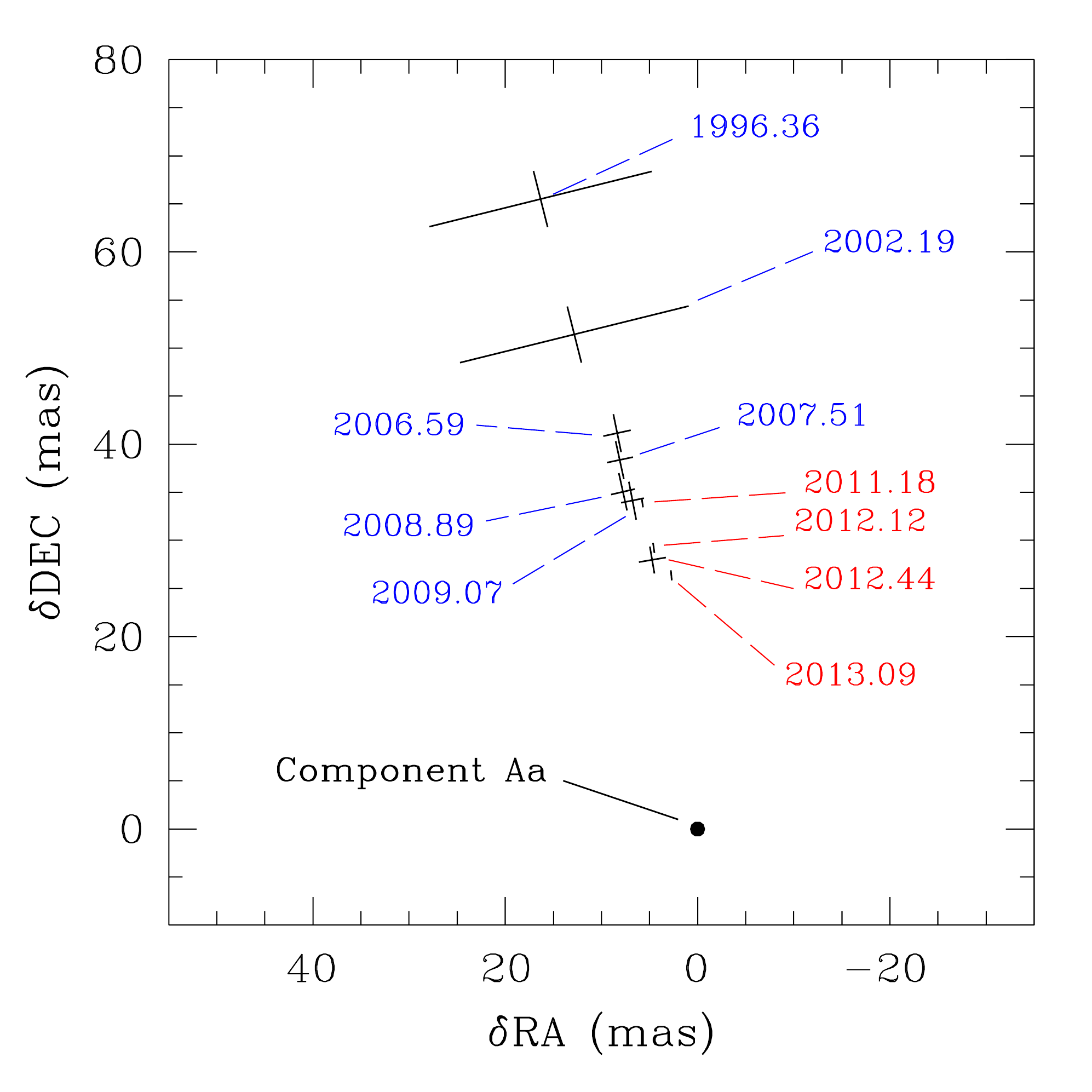}
      \caption{The relative positions, including error bars, of the
      two components of HD\,93129A, re-derived from archive HST/FGS
      (blue labels) data, along with VLT/NACO
      and VLTI/PIONIER (red labels: taken from 2011.18 to 2013.09) data \citep{Sana2014}. The epoch
      of the measurement is given for each point.  }
         \label{separation}
   \end{figure}

The relative motion of the two stars can be traced by archived HST/FGS
and HST/ACS\footnote{Hubble Space Telescope/Advanced Camera for
Surveys.} observations from 1996 to 2009, along with recent
VLT/NACO and VLTI/PIONIER\footnote{Very Large Telescope NAos-COnica
and Precision Integrated-Optics Near-infrared Imaging ExpeRiment.}
observations of Sana et al. (2014). Figure~\ref{separation} shows the
relative astrometry from these measurements over time.  The position
angles for 1996 and 2002 have large uncertainties because the binary
components were resolved along only one FGS axis, due to the $HST$
roll angle resulting in a small projected separation along the
unresolved axis. For later epochs, the approximate position angle was
known, so the observations were planned to avoid this problem. 

HD\,93129A was included in the recent Galactic O star Spectroscopic
Survey of \citet{Sota2014}. The coordinates from the Tycho catalogue
\citep{Hog2000} assume a single star since the presence of a companion
star was unknown at the time. Hipparcos did not resolved the
system. These coordinates have errors of about 0.1'', an order of
magnitude larger than the expected angular resolution of the LBA data
of $\sim10$~mas.

  \begin{table}[h!]
   \begin{center}
      \caption[]{Adopted parameters of HD\,93129A, relevant here.}
         \label{datos-hd93}
        \begin{tabular}{l r l c}
            \hline
            Parameter   & Value & Unit & Ref. \\
            \hline
            Aa Spectral type & O2 If+& & Wal02\\
   	   Ab Spectral type & O3.5 V& & Wal06 \\
            $\Delta m$   & 0.9$\pm$0.05 & mag & Nel10 \\
            Wind terminal velocity$\dag$  & 3200$\pm$200 & km s$^{-1}$ & Tar97 \\
            System mass & $200\pm45$ & M$_\odot$ & Mai08\\
            Effective temperature$\dag$ & 42500 & K & Rep04 \\
            System luminosity (log) & 6.2 & $L$/L$_\odot$ & Rep04 \\
            X-ray luminosity$\dag\dag$ &  1.3 x $10^{-7}$ & $L_{\rm bol}$ & Coh11 \\
            Distance & 2.5 & kpc & Wal95 \\
            \hline
     \multicolumn{4}{l}{$\dag$: Derived if a single star. $\dag\dag$: Derived from the energy}\\
  \multicolumn{4}{l}{ band (0.5 -- 8) keV.}\\ 
\end{tabular}
       \end{center}
       \tablebib{ 
	 Wal02: \citet{Walborn2002}; Wal06: Walborn, priv. communication; 
	 Nel10: \citet{Nelan2010}; Tar97: \cite{Taresch1997}; Mai08: \cite{Maiz2008};
	 Rep04: \citet{Repolust2004}; Coh11: \citet{Cohen2011}; Wal95: \citet{Walborn1995}.
       }
   \end{table}

  \begin{table*}
   \begin{center}
      \caption[]{PPMXL and FGS astrometry of the HD\,93129A field.}
         \label{astrometry}
        \begin{tabular}{l@{~~~}|@{~~~}c@{~~~}c@{~~~}c@{~~~}c@{~~~}|@{~~~}c@{~~~}c@{~~~}c@{~~~}c}
	    \hline 
	      & &  PPMXL & & & & FGS & \\
            Star &  RA (h) &  Dec (deg)  & $\mu_\alpha$ & $\mu_\delta$ & $\xi$ & $\eta$ & $\mu_\alpha$ & $\mu_\delta$ \\
\hline
  R1 & 10.7325347 (13) & -59.535576 (19) & -6.6 (2.5)  & -0.6 (2.5)  & -2.6531 (0.9) &  43.2955 (0.5) & -5.78 (0.29) & 1.54 (0.31) \\
   R2* & 10.7322835 (14) & -59.541006 (16) & ----  & ---- & -9.5351 (1.0) &  23.4780 (0.5) & -5.78 (0.53) & 1.71 (0.50) \\
   R3 & 10.7340882 (13) & -59.541208 (19) & -4.8 (3.2)  & 0.6 (3.2)   & 39.9186 (1.2) &  23.1199 (0.5) & -6.73 (0.32) & 1.38 (0.36) \\
   R4 & 10.7344007 (49) & -59.552781 (74) & 24.6 (17) & 60.5 (17) & 48.0711 (1.1) & -19.2493 (0.5) & -5.34 (0.44) & 2.22 (0.45) \\ 
   R6 & 10.7326563 (13) & -59.560735 (19) & -4.8 (2.1)  & 3.0 (2.1)  &   0.8902 (1.2) & -47.2647 (0.4) & -6.29 (0.32) & 1.64 (0.34) \\
   R7 & 10.7315615 (13) & -59.557925 (19) & -6.6 (2.5)  & 1.6 (2.5)  & -29.0924 (1.1) & -37.1969 (0.2) & -5.46 (0.29) & 1.60 (0.30) \\
   R8 & 10.7296359 (15) & -59.548569 (15) & -6.4 (2.0)  & 2.5 (2.0)  & -81.8936 (1.6) &  -3.6759 (0.3) & -5.53 (0.32) & 2.12 (0.30) \\
  \hline
 \multicolumn{9}{l}{ `Star': Reference star identification, same as
   in Figure \ref{fieldofhd}. All errors are shown within
   parenthesis. PPMXL coordinates}\\ 
\multicolumn{9}{l}{(J2000): RA in hours, Dec in degrees, with errors in
  mas. The J2000 FGS tangential plane coordinates, $\xi$, $\eta$, are
  given in}\\ 
\multicolumn{9}{l}{arcsec, relative to the point mid-way between the binary
  components; errors are in mas. Proper motions ($\mu$) in mas yr$^{-1}$,} \\ 
\multicolumn{9}{l}{ with errors
  in the same unit. *: Reference star 2 is not in the PPMXL catalog,
  so coordinates from UCAC4 are quoted.}\\ 
\multicolumn{9}{l}{The FGS proper motions are
  based upon FGS data but are constrained by the PPMXL
  values. The entries in this table }\\ 
\multicolumn{9}{l}{serve as the input
  values for the tangent plane analysis outlined in Section
  \ref{HST}.}
       \end{tabular}
   \end{center}
   \end{table*}

\section{Hubble Space Telescope/FGS data analysis}\label{HST}

\subsection{Astrometry}

 To improve upon the coordinates of HD 93129A from the Tycho
catalogue, we re-analyzed HST/FGS measurements.  The three Fine
Guidance Sensors on the $HST$ are 2-channel (x,y) white-light shearing
interferometers that provide line-of-sight positional measurements of
guide stars to enable accurate pointing and stabilization of the
$HST$. The FGS is capable of resolving sources down to $\sim$15 mas when
operated in its high angular TRANS  (or Transfer) mode and
sub-millarcsecond astrometry when operated in POS (or Position)
mode. A description of the FGS and its operation can be found in
\citet{Nelan2014}.

\begin{figure}
   \centering
  \includegraphics[width=0.5\textwidth]{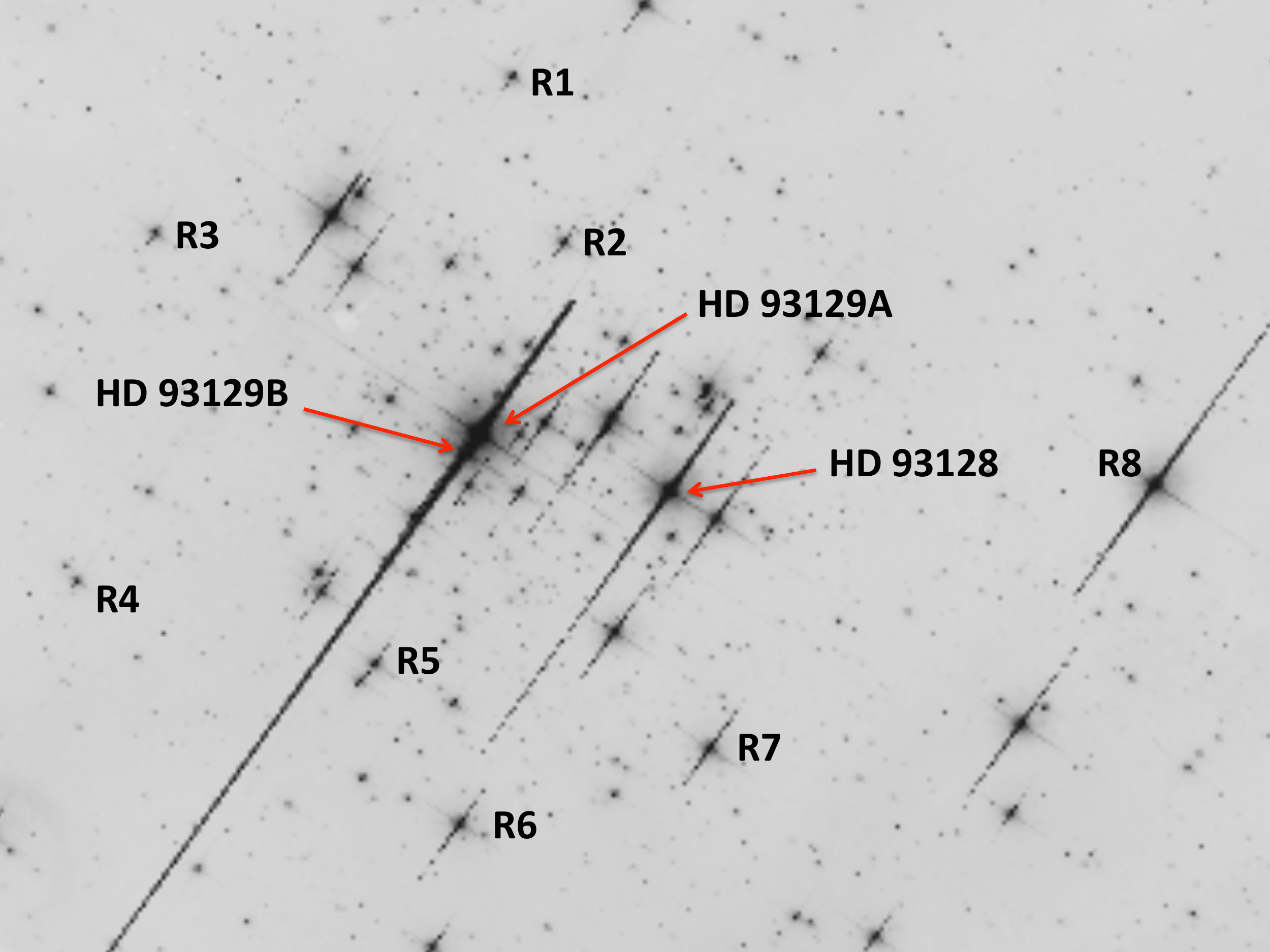}
      \caption{The core of Tr 14 and the selected FGS reference stars
      used in the astrometric measurements. The image (in log stretch)
      is from HST proposal 10602 using ACS/WFC with F550M, which is
      similar to the central pass band of the FGS measurements. The
      two stars HD\,93129A and HD\,93129B, separated by 2.7”, are not
      clearly resolved in this image since the roll angle of $HST$ at
      the time caused their position angle to be aligned with the
      diffraction spike. In this image the components of HD\,93129A
      are not resolved. Note that the separation between HD\,93129A
      and HD\,93128 is about 24”. North is up and East is to the left.}
         \label{fieldofhd}
   \end{figure}

In eight epochs between 2006 and 2009, FGS observed HD\,93129A in TRANS
mode, along with ``wide angle'' POS mode observations of the binary
itself and seven reference field stars within one arcminute (see
Fig.~\ref{fieldofhd}). The TRANS mode observations resolve
HD\,93129Aab, providing the component separation, position angle, and
magnitude difference at each epoch. The POS mode observations supply
the relative positions of the field stars and composite HD\,93129A
system. Combining data from these two observing modes gives the
position of Aa and Ab relative to these field stars at each
epoch. Combining all epochs yields a local FGS catalog of the relative
positions and proper motions of reference stars and binary components.

\begin{figure}
   \centering
  \includegraphics[width=0.45\textwidth]{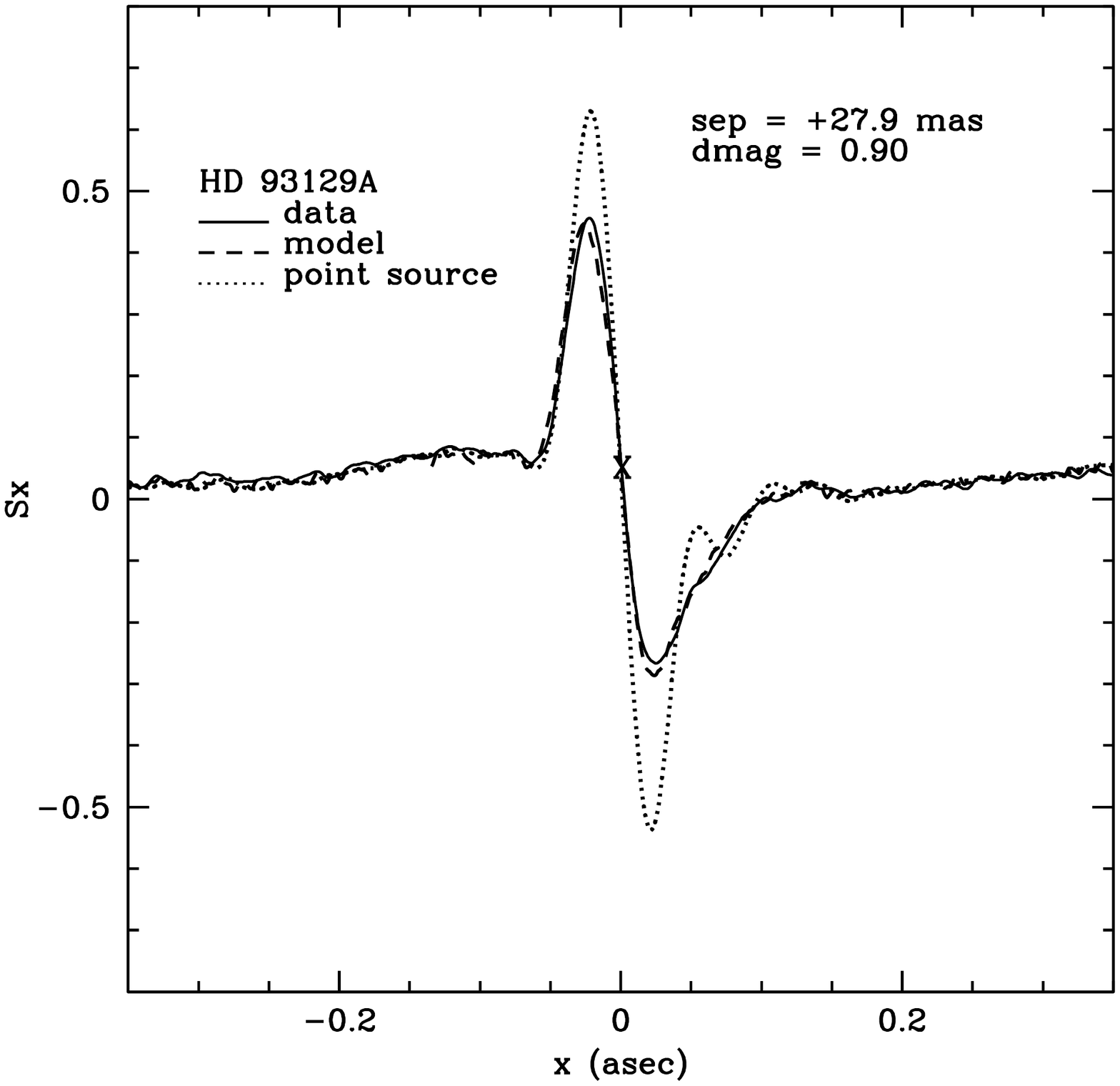}
   \includegraphics[width=0.45\textwidth]{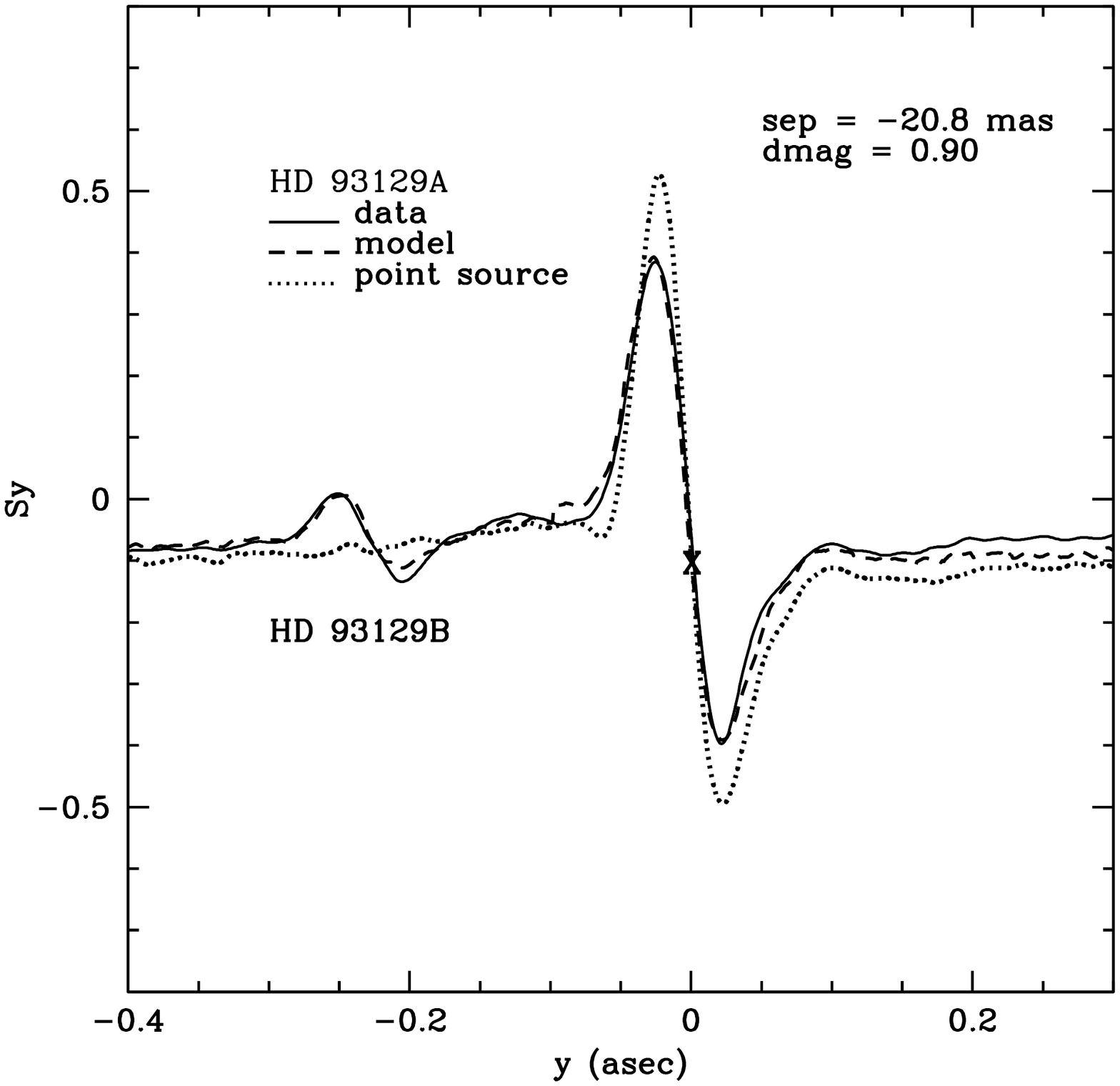}
      \caption{Interference fringes of HD\,93129A for Jan 27, 2009 FGS
      observations (solid line), the best fitting binary star model (dashed line) and, for
      comparison, the fringes of an unresolved point source
      (dotted line). The "X" shows the location of the composite fringe
      that POS mode observations tracked. Note that due to the roll of
      {\it HST} at the time of the observations, HD\,93129B has a
      projected separation from A along the y-axis of only 0.233",
      although its true distance is 2.73". }
         \label{interferencefringes}
   \end{figure}

Figure~\ref{interferencefringes} shows the interference fringes of
HD\,93129A for the Jan 27, 2009 observations, the best fitting binary
star model and, for comparison, the fringes of an unresolved point
source. The binary model takes into account the light contributed by
HD\,93129B which, 2.73'' away, accounts for about 5 percent of
 the light of HD\,93129A (FGS optics collimates a source onto
a circular beam with a full width-half maximum intensity diameter of
about 3''). Figure~\ref{interferencefringes} also shows the location
on the fringes that the POS mode observations of HD\,93129A ``lock'' on
to. The model fit to the TRANS mode observations provides the offset
of each component from this point, which thereby allows us to create
pseudo POS mode exposures for both HD\,93129Aa and HD\,93129Ab, and
hence accurate positions for each star relative to the reference
stars. The bright O star HD\,202124 was used as the point-source
calibrator since it is nearly the same B-V color as HD\,93129A, and it
was observed by FGS only eight days after HD\,93129A. The orientation
of {\it HST} at the time of the observation caused the projected
separation of HD\,93129A and B along the FGS y-axis to be only 0.233",
as is evident in the figure.

To compare the FGS coordinates of HD\,93129Aab to the radio source, it
is necessary to convert the relative FGS positions to absolute
International Celestial Reference System (ICRS) coordinates. The
FGS coordinate frame lies in a tangential plane, with the zero point
arbitrarily set to the point mid way between the binary components (at
the zero point the FGS Cartesian plane is tangential to the celestial
sphere). We used the PPMXL catalog\footnote{PPMXL is a catalog
of positions, proper motions, 2MASS and optical photometry of 900
million stars and galaxies.} \citep{Roeser2010} for the coordinates
and proper motions of the reference stars as input to a model that
 finds the RA,Dec of this tangent point. This is a two step
process. The FGS observations were obtained between 2006 and 2009,
while the PPXML coordinates are epoch J2000. The proper motions of the
reference stars must be taken into account for the most accurate
correlation of the FGS data with the radio observations. However, the
proper motions provided by PPMXL have large errors. As is standard in
FGS astrometry (e.g. Nelan \& Bond 2013, Benedict et al. 2007), the
PPMXL proper motions are input as ``observations with errors'' into
the model that combines the FGS astrometry from the eight epochs on to
the master plate. The model, which applies a 4-parameter plate
solution, outputs the best fitting proper motion for each reference
star, based upon the FGS measurements but constrained by the PPMXL
input values. The proper motions for HD\,93129Aa and HD\,93129Ab were
solved for directly without any external constraints. 
Table~\ref{astrometry} lists the reference star identifier, PPMXL 
coordinates and proper motions with the catalog errors, and the 
resulting FGS tangent plane coordinates and proper
motions and errors. The positional errors for RA, Dec are in
mas, while the proper motion errors are in mas yr$^{-1}$. The FGS
$\xi,\eta$ coordinates are in arcsecs, and their errors are in
mas. The FGS proper motions are now absolute rather than
relative. We note that PPMXL does not contain reference star
R2. For this star we used the UCAC4 ( Fourth US Naval
Observatory CCD Astrograph Catalog) coordinates but solved for the
proper motion using only FGS data.

Next step is to find the RA,Dec of the zero point of the FGS
tangential plane. The accurate FGS proper motions are used to update
the PPMXL coordinates to epoch 2009, which is the epoch of the FGS
master plate.  Using the algorithms outlined in \citet{Smart1960}, (1)
initial values for the RA,Dec of the FGS tangent point are given, for
which the Guide Star Catalog v2.3 coordinates of HD\,93129A are
used, (2) the PPXML (RA,Dec) of each reference star are converted to
Cartesian coordinates in this tangential plane, which are treated as
``observations with errors'', (3) the PPXML Cartesian coordinates are
used to set the scale and orientation of the FGS frame to
(East, North). (4) An iterative process is applied to find a global
solution that provides the best-fit of the observed FGS and PPMXL
coordinates in the tangential plane. The solution creates a hybrid
FGS-PPMXL catalog and computes the RA,Dec of the tangent point. Once
convergence is achieved, it is a straightforward process to compute
the (RA,Dec) of any point in the FGS frame. We used the GAUSSFIT
\citep{Jefferys1987} least-squares program to build the models to
carry out these calculations.

We compute the RA,Dec of the HD\,93129Aa and Ab components,
adjusted for proper motion to the Aug 6, 2008 epoch of the radio
observations, equinox J2000:
\begin{eqnarray}
\alpha, \delta ({\rm Aa}) = 10^{\rm h} 43^{\rm m} 57.455^{\rm s}, -59^\circ 32' 51.36'', \nonumber \\
\alpha, \delta ({\rm Ab}) = 10^{\rm h} 43^{\rm m} 57.456^{\rm s}, -59^\circ 32' 51.33''. \nonumber 
\end{eqnarray}
The uncertainty in each coordinate is $\pm27$ mas, dominated by the
PPMXL catalog errors in the position and proper motion of the seven
FGS reference stars surrounding HD\,93129A. However the position of Aa
relative to Ab is accurate to about 1 mas. From these positions, the
separation and position angle of Aab in Aug 2008 was $36\pm 1$ mas and
$PA = 12\pm 1^{\circ} $.

The proper motions derived from the FGS data are $\mu_\alpha,
\mu_\delta ({\rm Aa})$ = $-8.4, 2.6\,\, {\rm mas\, yr^{-1}}$ and
$\mu_\alpha, \mu_\delta ({\rm Ab})$ = $-9.0, 0.0\,\, {\rm mas\,
yr^{-1}}$.  Note that the proper motion of the binary's components
include any putative orbital motion along with the system motion. The
difference from the reference star proper motions
(Table~\ref{astrometry}), particularly in RA, could reflect internal
motion within the core of Tr 14, while the reference stars lie outside
the core.

\subsection{Orbit estimation}

   \begin{figure}
   \centering
\includegraphics[width=9cm,angle=0]{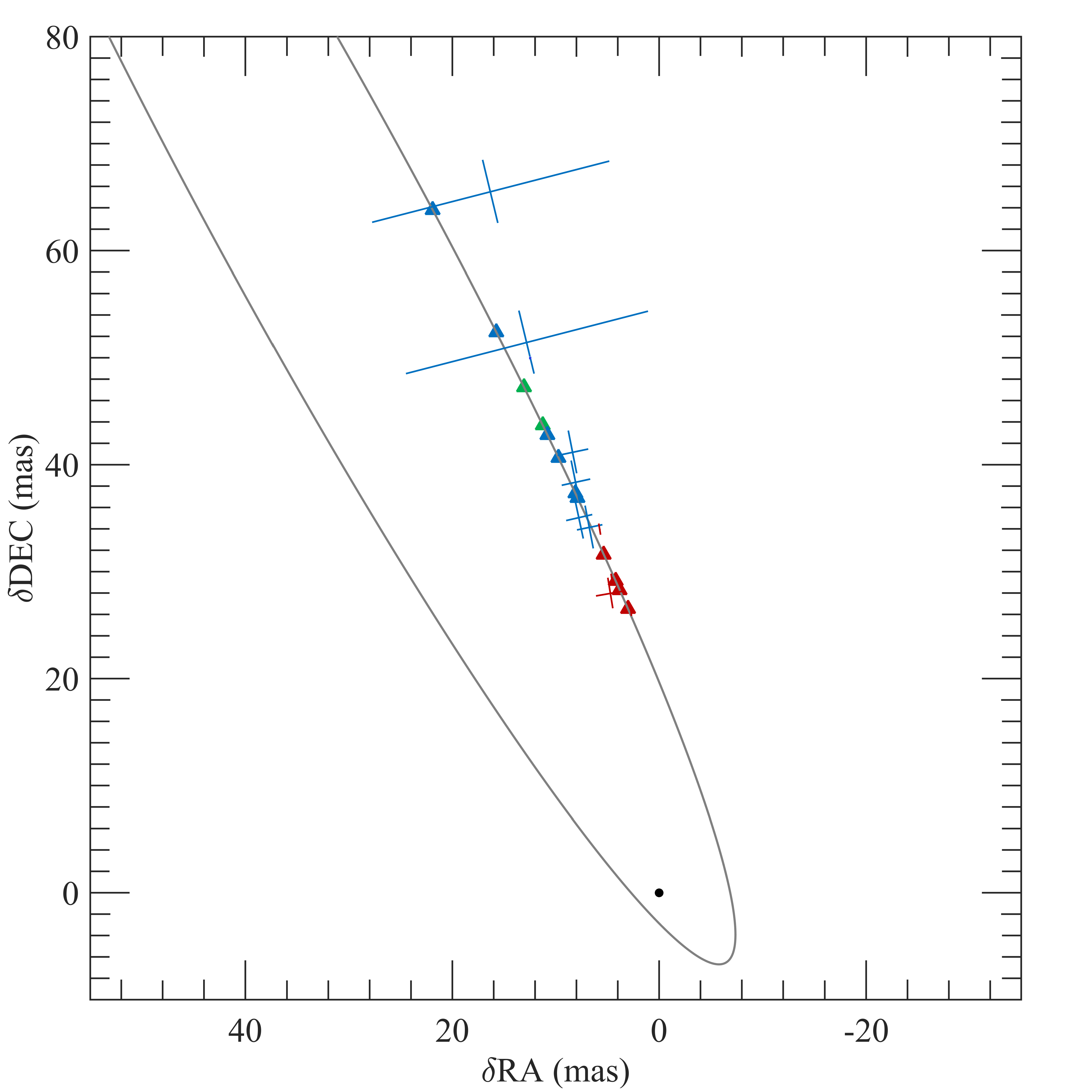}
\caption{A preliminary fit for the orbit (triangles). The primary Aa
  is at the origin, The different positions of the secondary Ab
  between 1996 and 2013 are represented with crosses, and are the same
  as in Fig. \ref{separation}.  The figure shows the projection of the
  preliminary orbit in the plane of the sky.  }
         \label{orbitalfit}
   \end{figure}

 The relative motion of the components Aa and Ab hints at HD\,93129A 
being a gravitationally bound  system. However, the relative positions 
shown in Figure 1 cover a rather small part of the orbit and thus are
inadequate to perform a standard determination of orbital
elements. No definitive radial velocity variation has been
reported, which is not surprising given the wide separation and
apparent near linear trajectory of the stars. Nevertheless, a
first approximation of an orbital fit was attempted. An algorithm that
minimizes the weighted square distance between the measured data
points and the fitted orbit (Casco \& Vila, private communication) was
used to determine a preliminary set of orbital parameters. Our results
suggest an orbital period of the order of 200 years, an eccentricity
larger than 0.9, and semi-major axes of about 93 and 37 mas
respectively for components a and b. Our best-fit solution 
points also to a periapsis argument of about 220 degrees and an
inclination of about 103 degrees. An attempt to fix the inclination
value close to zero yielded a much poorer fit (larger weighted root
mean square rms). According to this preliminary solution, the next
periastron passage is expected to take place in 2024. It should be
emphasized that such a preliminary solution was calculated to give a
rough idea of the orbit, even though the present astrometry does not
allow for a convincing characterization of the orbit accompanied by
relevant error bars on the estimated parameters. This calculation has
however the merit to provide a preliminary basis to organize future
observation campaigns dedicated to HD 93129A.

\section{Radio observations}
\label{lba}

\subsection{Observations of June 22, 2007}\label{obs2007}

We observed HD\,93129A on June 22, 2007 (MJD~54273) between 03:00 and
13:00~UTC at 2.3~GHz as part of eVLBI experiment vt11D3 that used
Parkes, Mopra, and Australia Telescope Compact Array (ATCA) to
search for suitable phase calibrators near the target
\citep{Benaglia2010}. The sources PKS 0637$-$752 (J0635$-$7516) and
J1047$-$6217 were used as flux and phase calibrators,
respectively. The total time observing HD\,93129A was 3 hours.

\begin{figure}
   \centering
  \includegraphics[width=0.45\textwidth]{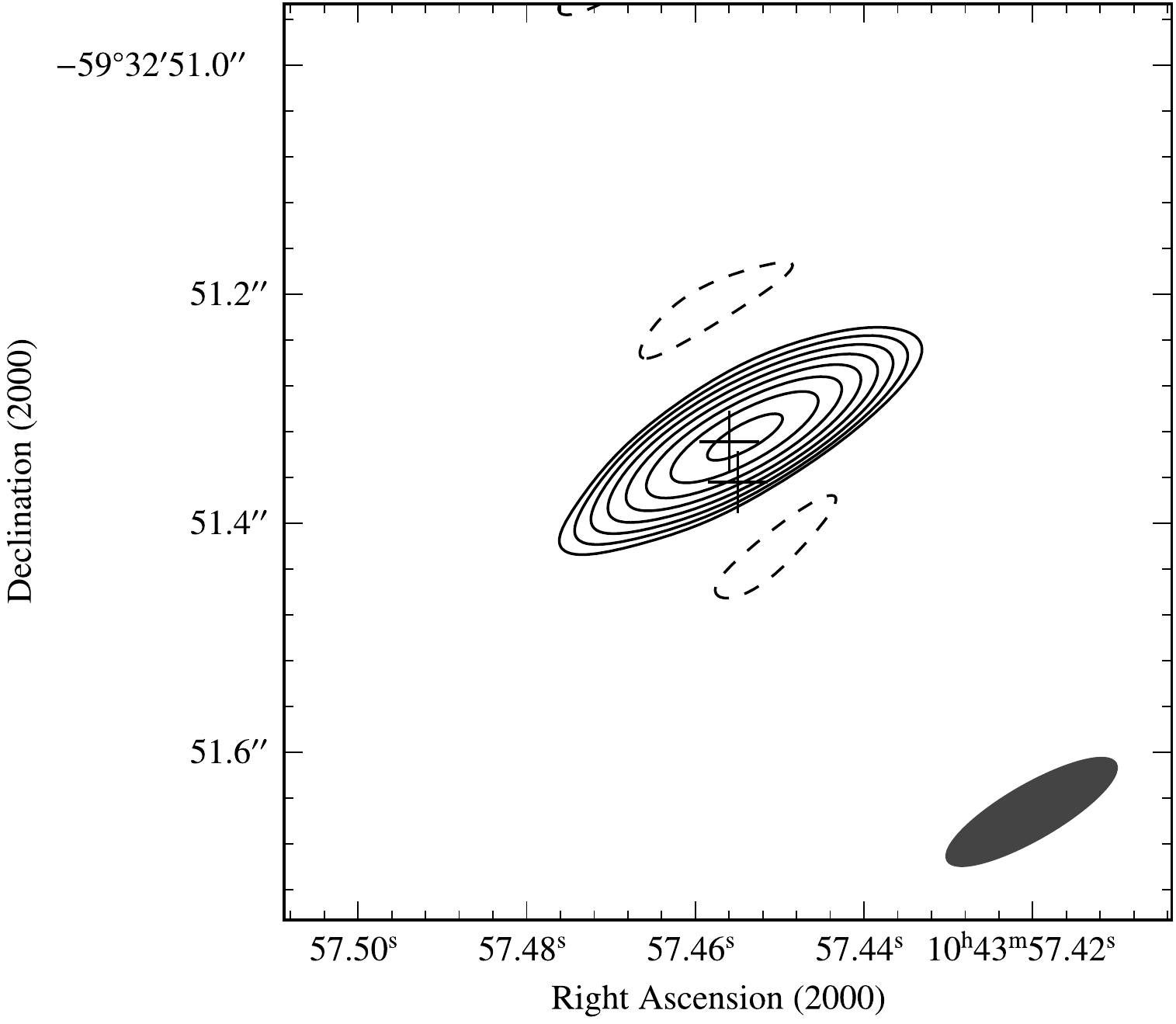}
\caption{Image of the HD\,93129Aab system, from 2007 LBA data (in
  contours) and the location of stars Aa (southern) and Ab
 (northern) marked with crosses.  The contours start at $0.3$~mJy
 beam$^{-1}$ ($3\sigma$) and increase by factors of $2^{1/2}$. The
 synthesized beam ($0.2'' \times 0.05''$, $PA = 31^{\circ}$) is shown
 at the bottom right corner. North is up and East is to the left.}
         \label{i2007-8}
   \end{figure}

Standard VLBI calibration procedures were used (see also
Sect.~\ref{obs2008}) and the source imaged with a synthesized beam of
$0.2'' \times 0.05''$, $PA = 30.9^{\circ}$. The source has a measured
flux density of $\sim 3$ mJy. See more details in \cite{Benaglia2010}
and Fig.\,\ref{i2007-8}.

\subsection{Observations of August 6, 2008}\label{obs2008}

\begin{figure}
   \centering
  \includegraphics[width=7cm,angle=0]{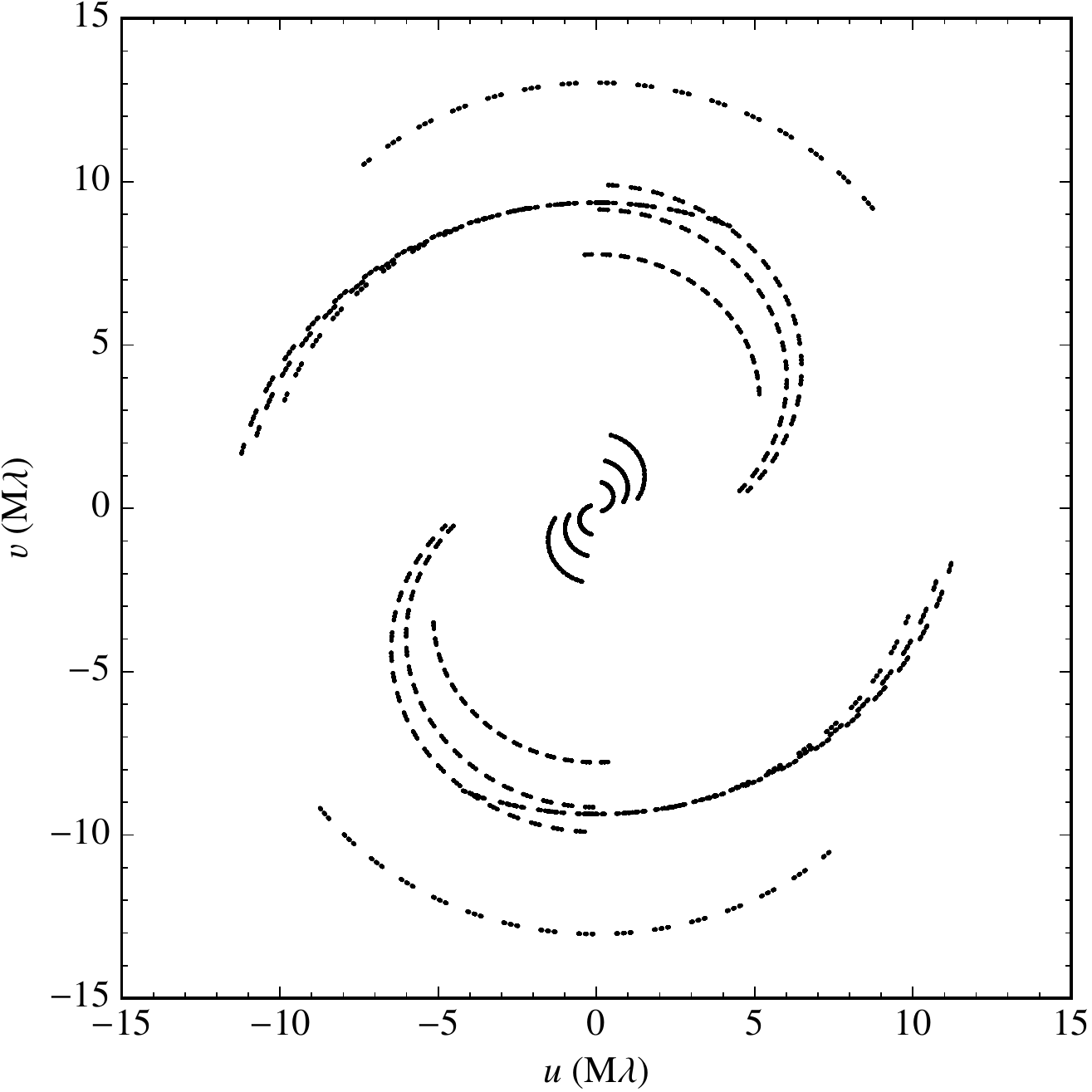}
      \caption{The $uv$ coverage of project V191B in Aug 2008.}
         \label{uvcoverage}
   \end{figure}

Further observations were conducted on Aug 6, 2008 (MJD~54684) from
01:00 to 12:00~UTC with the LBA at an observing frequency of 2.3 GHz,
and using five antennas: ATCA (in tied-array mode, i.e., working
as a single station), Parkes, Mopra, Ceduna, and Hobart (experiment
V191B). The baseline lengths range from 100 to 1700 km, providing an
angular resolution of $\sim15$~mas at that frequency. The data
were recorded at 256 Mbps provided by 4 sub-bands of 16-MHz
bandwidth at each of the two circular polarizations. The data were
correlated with the Distributed FX (DiFX) software correlator
located at Curtin University of Technology (WA, Australia), using 128
channels per sub-band, each band 125 kHz wide, and using 2-second
integrations.  The data from the ATCA antennas were also
correlated as an independent interferometer to obtain the total flux
density from a low-resolution observation. The time on HD\,93129A
amounted to 3.2 hours. The $uv$ coverage is shown in
Fig.\,\ref{uvcoverage}.

The data reduction was performed using the Astronomical Image
Processing System (AIPS)\footnote{\url{http://www.aips.nrao.edu}}
software package. Bad data were flagged using telescope a-priori flag
files and information provided in the observing logs. The amplitude
calibration was performed using the antenna system temperatures
and phase solutions on the calibrators were obtained using the
AIPS task {\sc fring}. The compact source J1047$-$6217
was used as the phase calibrator.  The fringe finders 0637$-$752
(J0635$-$7516), 1549$-$790, J1051$-$6518, J1023$-$6646 and the phase
calibrator J1047$-$6217 were used for the bandpass calibration.
First, an accurate model of the calibrator was obtained after several
self-calibration and imaging cycles. The calibrator, slightly resolved
on the longest baselines, had a flux density of $0.99\pm0.02$~Jy.
Finally, the phase solutions and the amplitude scale were transferred
to the target source, and images produced by deconvolving the
visibility data.

The data from ATCA were correlated in standard interferometer
mode. The resulting fluxes for the phase calibrator J1047-6217 and
HD\,93129A were 1.300$\pm$0.002 Jy and 7.5$\pm$0.1 mJy
respectively. The flux and uncertainty for the phase calibrator was
derived by fitting a point source to the image, using all baselines.

\subsection{ATCA observations}

 The Australia Telescope Online Archive (ATOA) was searched for 
observations of HD 93129A. In addition to project C678 (Benaglia et
al. 2006) and the observations associated with the LBA projects
described in the previous Section, project C1726 targeted HD 93129A
for 3.3 h at 4.8 and 8.6 GHz. Table~\ref{atca-fluxes} summarizes the
dates and observing frequencies for these observations.  In all
observations with the ATCA, HD\,93129A is not resolved by ATCA and
appears as a point source.

\begin{table}
   \begin{center}
      \caption{Radio flux density of HD~93129A at different
        frequencies obtained with ATCA and LBA data. We include the
        project code and the flux density calibrator name with its
        flux value.}
        \begin{tabular}{l r r l r}       
            \hline
            Array/project  & $\nu$ & S$_\nu$ & Flux cal& $S_{\rm \nu, cal}$ \\
             and Date    & [GHz] &   [mJy]     &  name & [Jy]\\
            \hline
            ATCA/C678 &&&&\\
                  $\; \; \;$ 2003-01-28  & 4.8 & 4.1 $\pm$ 0.4$^{\rm a}$ & 1934$-638$ &2.84\\
                  $\; \; \;$ 2003-01-28  & 8.6 & 2.0 $\pm$ 0.2$^{\rm a}$ & 1934$-638$ & 5.83 \\
                  $\; \; \;$ 2003-12-20  & 1.4 & 9.4 $\pm$ 0.9$^{\rm a}$ &1934$-638$ &14.98 \\
                  $\; \; \;$ 2003-12-20  & 2.4 & 7.8 $\pm$ 0.4$^{\rm a}$ &1934$-638$ &11.59 \\
	         $\; \; \;$ 2004-05-05  & 17.8 & 1.8 $\pm$ 0.15$^{\rm a}$ & Mars & \\
		$\; \; \;$ 2004-05-05  & 24.5 & 1.5 $\pm$ 0.35$^{\rm a}$ & Mars & \\
	  ATCA/V191B &&&&\\
	        $\; \; \;$ 2008-08-06 & 2.3 & 7.5$\pm$0.11 &0637$-$752 &  5.32 \\
	 ATCA/C1726  &&&&\\
	        $\; \; \;$ 2009-01-18 & 4.8  & 5.6$\pm$0.3 & 1934$-638$ & 5.83\\
	        $\; \; \;$ 2009-01-18  & 8.6 & 2.9$\pm$0.3 & 1934$-638$ &2.86\\
            LBA/V191B  &&&&\\
                  $\; \; \;$ 2008-08-06  & 2.3 & 2.9$\pm$0.51$^{\rm b}$ & 0637$-$752 &  \\	  
            \hline
       \end{tabular}
\tablebib{a: Benaglia et al. 2006; b: Derived from image portrayed in Fig.~\ref{radioimage}-right.}
\label{atca-fluxes}
   \end{center}
\end{table}

\section{Results}
\label{results}

\subsection {Radio flux density}
 
In Fig.~\ref{radioimage} we show the radio image from LBA 2008
observations of HD\,93129A at 2.3 GHz. Two final images are presented:
one at maximum angular resolution (Fig.~\ref{radioimage}-left:
high-resolution image), and a second one using tapering
that increases the sensitivity to larger scale emission 
(Fig.~\ref{radioimage}-right: low-resolution image).  The
high-resolution image has an rms of 0.21~mJy~beam$^{-1}$, and shows a
``comma-shaped'' extended emission with a flux density of $1.5\pm
0.5$~mJy. The value quoted for the error is the difference in the
integrated fluxes above 2 and 3$\sigma$, where $\sigma$ is the
measured rms of the image.  In the low-resolution image the rms noise
is lower and more flux is recovered from the source, as expected. The
rms of the image is 0.13~mJy~beam$^{-1}$, and a gaussian fit to the
source gives an integrated flux density detected by the LBA of
$2.9\pm0.5$~mJy, with a peak intensity of $1.8\pm0.2$~mJy
~beam$^{-1}$.

We note that the amplitude calibration of the data, which is based on
the system temperature of the antennas is limited by the
uncertainty on the antennas' response and gain curves. The flux
density measured for the phase calibrator is $\sim30$\% lower than its
unresolved flux density measured with ATCA between Feb 2007 and Nov
2008: $1.46\pm 0.01$ Jy to $1.42\pm 0.02$. This indicates that
calibrator is partially resolved by the LBA or it is variable, or both. 
We also note that the LBA-measured flux of HD\,93129A is
$\sim$35\% of the total flux measured in simultaneous Aug 2008 ATCA
data at 2.3~GHz. We conclude that the rest of the flux from HD\,93129A
is resolved out by the LBA.

\subsection{Radio morphology and astrometry} 

 From the 2008 LBA image the angle subtended by the emitting source 
as seen from the secondary star (so-called opening angle 2$\theta_{\rm w}$) 
can be estimated, leading to half-opening angle 
$\theta_{\rm w} \sim$ 65$^\circ$. This is related to the wind-momentum rate ratio
$\eta$ by $\theta_{\rm w} = 180^\circ \times \eta / ( 1 + \eta)$, thus implying
$\eta \sim 0.6$.

The radio emission at maximum angular resolution presents two
maxima. 
 The position of the centroid of the two fitted gaussians is 
$\alpha_{\rm C}$ = 10:43:57.456, $\delta_{\rm C}$ = $-$59:32:51.34, J2000.

\begin{figure*}
   \centering
   \includegraphics[width=9cm, angle=0]{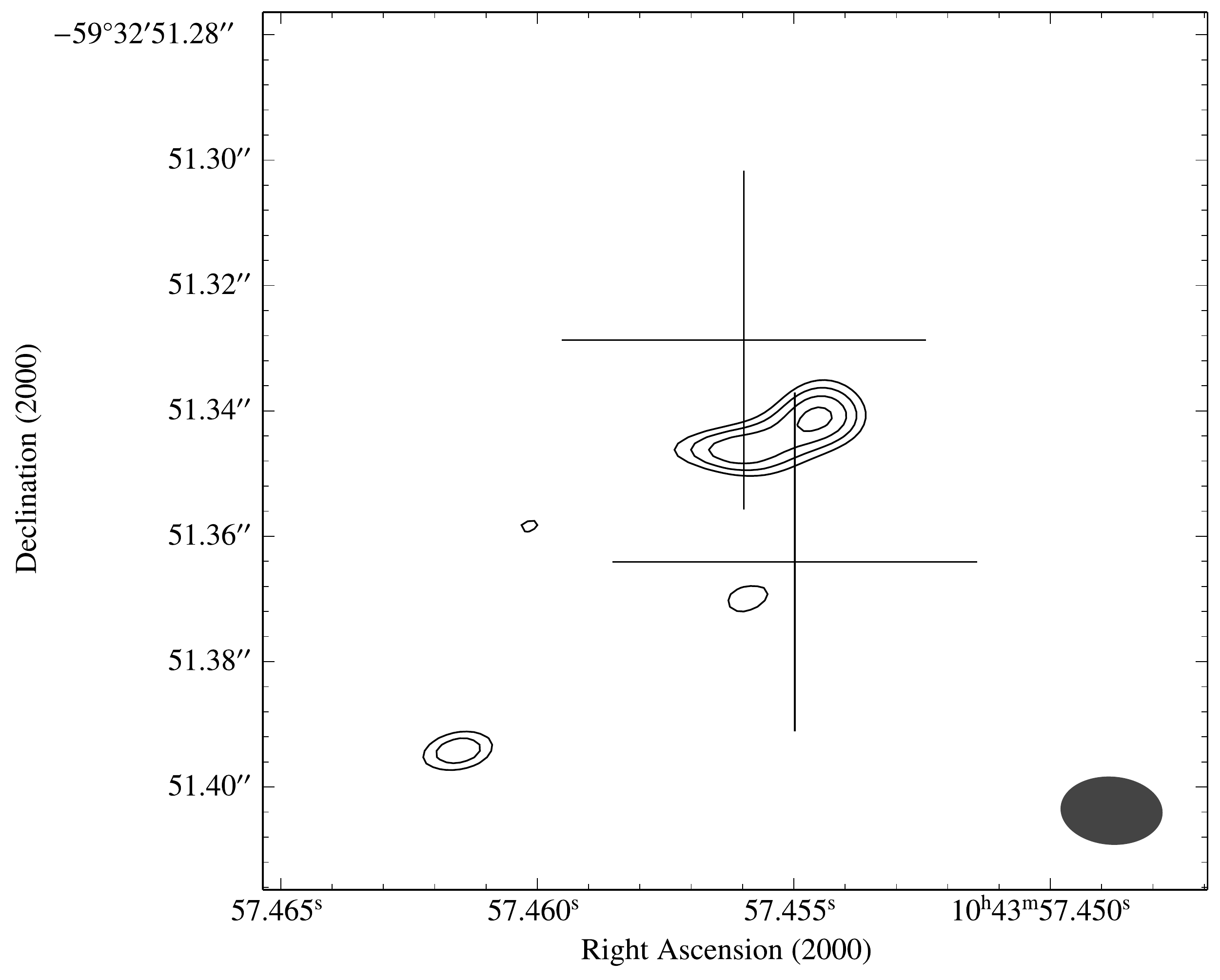}
   \includegraphics[width=9cm, angle=0]{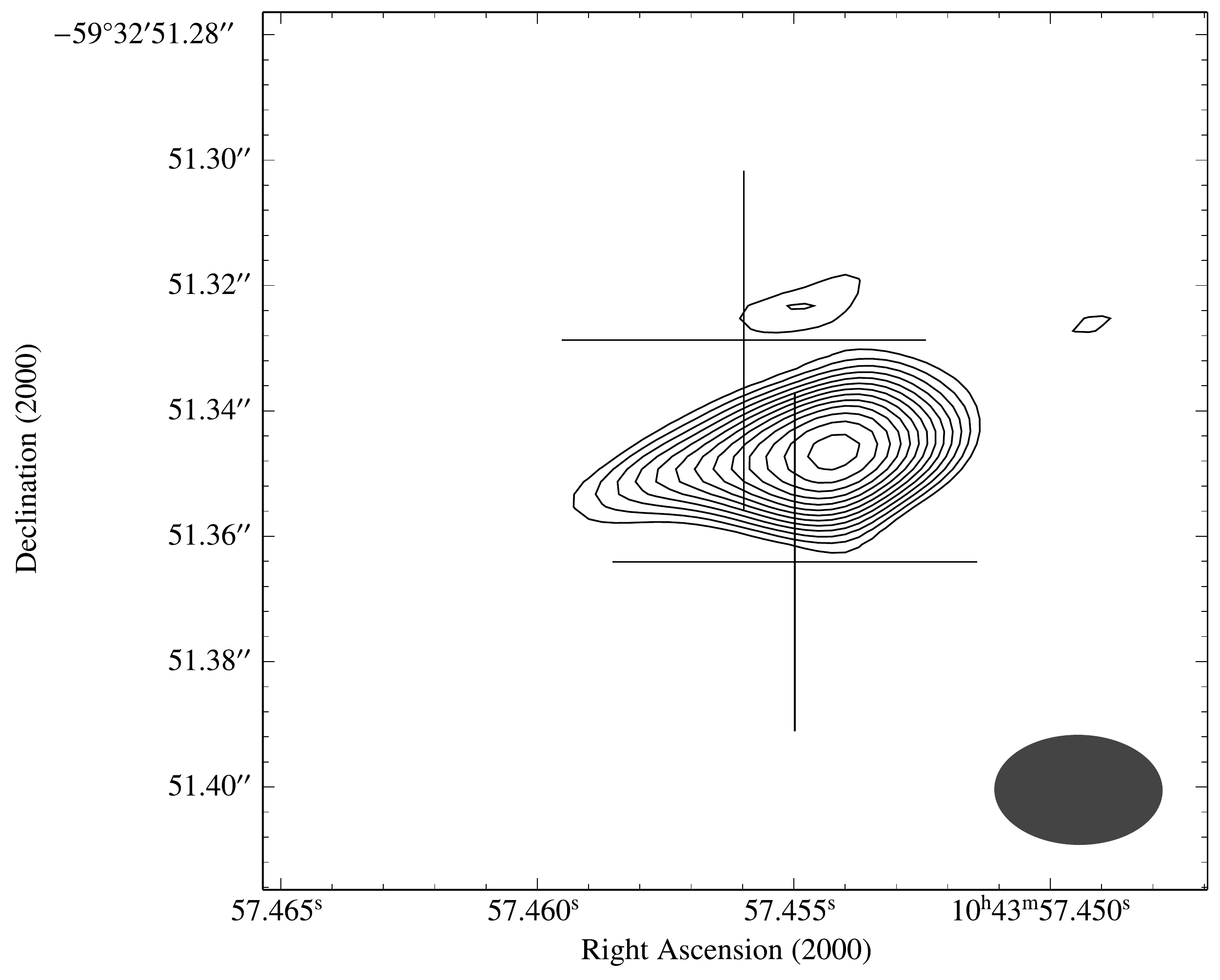}
   \caption{The LBA observations of HD\,93129A at 2.3 GHz. a) Left: At
     highest angular resolution with an image rms of 0.21~mJy
     beam$^{-1}$ and contour levels shown for -0.6, 0.6 (3$\sigma$),
     0.7, 0.8 and 0.9 mJy beam$^{-1}$. b) Right: a tapered image that
     increases the weight of the shorter baselines and increases
     sensitivity to larger scale emission; the image rms is 0.13~mJy
     beam$^{-1}$ and the contour levels are -0.4, 0.4 (3$\sigma$),
     0.6, 1, 1.4 and 1.8 mJy beam$^{-1}$. The synthesized beams are
     $15 {\rm mas} \times 11 {\rm mas}$ and $31 {\rm mas} \times 23
     {\rm mas}$ respectively, shown in the bottom right-hand corner of
     the images. The crosses mark the positions of the system
     components Aa (south) and Ab (north), at the epoch of the radio
     observations based on the position derived in Sec.~\ref{target}.
     North is up and East is to the left.}
\label{radioimage}
    \end{figure*}

A systematic uncertainty of 1.1 mas in Right Ascension and 2.8 mas in
Declination comes from the position error of the phase reference
calibrator in the ICRF2 (International Celestial Reference Frame,
second realization). Therefore, the absolute uncertainty of the
position of the radio emission is about 3~mas in the ICRF. In
Fig.~\ref{radioimage} the crosses mark the positions of HD\,93129 Aa
and Ab, with the size of the crosses indicating the absolute
uncertainty, although the relative separation of the components has an
uncertainty of $\sim1$~mas (see Sect.~\ref{HST}).

\section{Discussion}\label{discussion}

\subsection{Wind-momentum rates}
The radio source appears in the LBA observations as a bow-shaped
region, slightly curved around the component Ab, and
reminiscent of the WCRs resolved in WR\,140 \citep{Dougherty2005}, WR\,146
\citep{WCRwr146}, Cyg OB2 \#5 \citep{Ortiz2011}, as examples. If we
assume that the radio source is centered on the stagnation point of
the WCR, the wind momentum rates ratio $\eta$ can be expressed
as:
$$\eta =
\left(\frac{R_b}{R_a}\right)^2 = \frac{\dot{M}_b v_{b}}{\dot{M}_a
v_{a}},$$
\noindent where $R_i$ is the distance between the WCR and
star $i$ \citep{usov1992}. Note that the value of $\eta$ is independent
of orbit inclination.  

From the relative position of Aa, Ab and the centroid of the radio
emission in Figure 7, we estimate $R_b / R_a \approx 0.7$.  The
stellar winds of the two stars are assumed to have reached their
terminal velocities before colliding, considering the wide orbit, and
we further assume that the terminal velocities of both components are
similar (see the values quoted for instance in Table\,1 of Muijres et
al. 2012), then $\eta$ reduces to a mass-loss rate ratio, leading to
$\dot{M_b} \approx 0.5 \dot{M_a}$.

\begin{figure}
   \centering
  \includegraphics[width=0.45\textwidth]{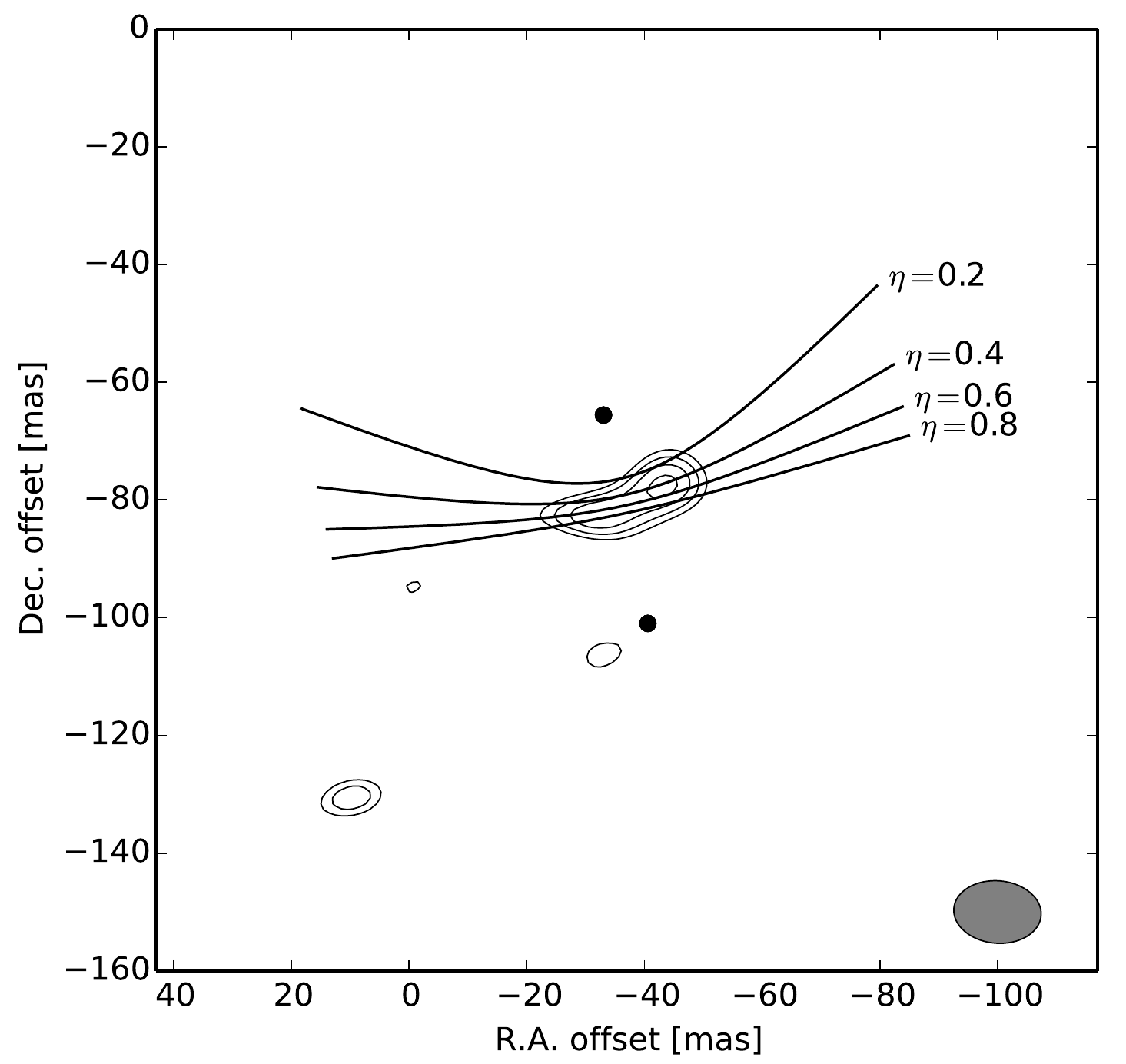}
\caption{Image of the HD\,93129Aab system at 2.3 GHz, showing different
$\eta$ values. The black dots mark the position of the stars. The
synthesized beam is shown in the bottom right corner.}
         \label{etas}
   \end{figure}

In this work, we hesitate to use the shape of the WCR to estimate
$\eta$. As \citet{PD2006} (section 4.3.3) show for WR\,140, the
emission in the WCR detected in VLBI is close to the stagnation point
and the opening angle has not reached the asymptotic value. They also
noted the challenge of identifying the stagnation point relative to
the stellar components due to opacity effects.  In the best case,
we can perhaps assume the shape of the contact discontinuity (CD)
between the two stellar winds \citep{canto1996} 
is represented
by the radio emission. In Figure \ref{etas}, the positions of the CD
corresponding to different values of $\eta$ are plotted relative to
the emission detected in the LBA observation. The distance and angle
between stars are known, so the only free parameters are the offsets
with respect to the radio image and $\eta$. The $\eta$ values that
best match the radio source are between $\sim 0.4$ and $\sim 0.6$.

\subsection{On the binary nature of HD\,93129A}

Do HD\,93129Aa and Ab form a gravitationally bound binary system? The
system is being monitored in the OWN Survey project
\citep{BarbaOwn}.  A radial velocity (RV) curve from preliminary data
shows slight variations though no period has yet been
determined. The putative low amplitude RV variations may be due to a
number of factors, including insufficient data over a potential
orbital period, a very low mass companion or unfavorable orbital
inclination (Barb\'a \& Gamen, private communication).

An argument in support of HD\,93129Aab being a bound system is the
small angular separation and the change in the rate at which the two
stars are approaching one another. In the simplest scenario, if they
are in the plane of the sky, with a separation of 26.5 mas (epoch
2013.09 from Sana et al. 2014) and at a distance of 2.5 kpc, they
have a physical separation of only 66 AU, which implies a bound system
for such massive stars. However, this is strongly dependent on the
inclination being in the plane of the sky.
More compelling is the increase in the rate at which the two stars are
approaching one another. Based upon FGS and ACS data, between 1996 and
2009 the separation had been closing at approximately 2.4 mas
yr$^{-1}$, but VLTI data shows this rate to have accelerated to 4.2
mas yr$^{-1}$ between 2011.19 and 2013.09 \citep{Sana2014}, consistent
with a binary system approaching periastron. The relative positions in
Fig.~\ref{separation} suggest either an orbital inclination
close to 90 degrees, or a nearly face-on orbit of high
eccentricity. The results of the very preliminary orbit fits given
in Sect. \ref{HST} favor the former hypothesis.

\subsection{On the mass-loss rates of the HD\,93129A system}

HD\,93129A has been the target of different studies aimed at deriving the
mass-loss rate ($\dot{M}$)\footnote{For an overview of these mass-loss
rate determination techniques, including a comprehensive discussion of
their advantages and limitations, we refer to the review by
\citet{PVN2008}.}. \citet{Taresch1997} estimated $\dot{M}$ of
HD\,93129A assuming it was a single star, by fitting the H$\alpha$
profile and also using a more complete spectral synthesis approach in
the ultraviolet. These two approaches led to consistent values of
1.8\,$\times$\,10$^{-5}$ and
2.1\,$\times$\,10$^{-5}$\,M$_\odot$\,yr$^{-1}$, respectively. The
authors considered that these values should be reduced by a factor
$2^{3/4}$, a reduction of about 40\%, if HD\,93129A consisted of two
very similar components. Using  H$\alpha$ line diagnostics exclusively,
Repolust et al. (2004) derived a value of
2.4\,$\times$\,10$^{-5}$\,M$_\odot$\,yr$^{-1}$ for the mass-loss
rate, noting that the results could strongly suffer from
contamination by a companion star.

\citet{BK2004} presented flux density measurements of the HD\,93129A
system from ATCA data at two frequencies, 4.8 and 8.64 GHz, and
derived a spectral index $\alpha = -1.2\pm0.3$. Their interpretation
was that both thermal and non-thermal emission were contributing to
the radio emission. \citet{WB75} showed that the mass-loss rate of a
star with a thermal wind can be expressed in terms of the radio flux
density. Using the \cite{WB75} formulae and if $f_{\rm T}$ is the
fraction of thermal emission at a given frequency, it is possible to
express the mass-loss rate in terms of $f_{\rm T}$ as $\dot{M} =
(f_{\rm T})^{3/4} \times 7.2 \times 10^{-5}$ M$_{\odot}$ yr$^{-1}$ at
$\nu = 8.64$~GHz for a stellar distance of 2.5 kpc. This leads to a
mass-loss rate of $\approx 5 \times 10^{-5}$ M$_{\odot}$ yr$^{-1}$ if
$f_{\rm T} = 0.5$.

At the epoch of the observations presented by \citet{BK2004}, the
stars Aa and Ab were $\sim$ 50 mas apart in the plane of the sky (see
Sect. 2).  The angular resolution of the ATCA data at $\sim1$ arcsec
precludes separation of the contributions to the total flux of both
stellar winds and a colliding-wind region. An estimate of the thermal
emission from the two stellar winds can be obtained from the radio
observations in \citet{Benaglia2006} from ATCA at 17.8 and 24.5
GHz. At these frequencies, non-thermal emission is negligible and can
be disregarded to zeroth order. The total flux is then the sum of the
thermal emission from the two stellar winds, and characterized with a
spectral index of $+0.6$ \citep{WB75}. Consequently, the thermal
contribution to the flux at lower frequencies can be derived by
extrapolation and assuming no variations between 2003 (8.64 GHz data)
and 2004 (17.8 GHz data). This leads to a thermal contribution at 8.64
GHz of 1.2 mJy, compared with a total flux of 2.00$\pm$0.15 mJy (see
Table~\ref{atca-fluxes}), an $f_{\rm T} \approx 0.6$, and $\dot{M}=
4.9 \times 10^{-5}$ M$_{\odot}$\,yr$^{-1}$.

\begin{table}
   \begin{center}
      \caption{Mass loss rates values derived for HD\,93129A.}
         \label{masslossrates}
        \begin{tabular}{l l r}       
            \hline
            Reference  & Method & Value $ \times 10^5$ \\
                                &                 & (M$_\odot$ yr$^{-1}$)\\
              \hline
            Tar97 & H$\alpha$ profile  &  1.8  \\    
            Tar97 & Ultraviolet lines      &  2.08 \\  
            Rep04& H$\alpha$ profile  &  2.36 \\
            BK04 & 8.64GHz continuum flux  &  $(f_{\rm T})^{3/4}$ 7.2$\dag$\\
            This work & Separating T and NT fluxes & 4.9$\dag\dag$ \\      
            \hline
       \end{tabular}
\tablebib{Tar97:  \citet{Taresch1997}; Rep04: \citet{Repolust2004}; BK04: \citet{BK2004}; $\dag$: $(f_{\rm T})$ $\epsilon$ [0,1]; $\dag\dag$: see text.}
 \end{center}
\end{table}

The main conclusion here is that mass-loss rate determinations are
challenging when dealing with a binary system. Considering the high
fraction of binaries among O-type stars \citep[see e.g.][and
references therein]{Sana2014}, many observational determinations of
${\dot M}$ should certainly be viewed with caution, including those of
HD\,93129A.

\subsection{Flux density variability of HD\,93129A}

 The flux density measurements of HD 93129A derived from ATCA 
data (Table~\ref{atca-fluxes}) cover frequencies from 1.4 to 24.5 GHz.
Observations at 2.4, 4.8 and 8.6 GHz were obtained twice, at different 
epochs. The two measurements at 2.4~GHz were observed in 2003 and 2008, and
the fluxes agree within the uncertainties.

The fluxes at 4.8 ~GHz (from 2003 and 2009) differ by
$\sim37$\%, increasing with time. Similarly, the 8.6~GHz fluxes at the
same epochs differ by $\sim45$\%. In all the archival
observations, relative errors are 10\% or less. We checked for
calibration errors in the individual datasets. For all of them, the
flux calibrator was 1934-638, from which the flux scale of phase
calibrators and source were bootstrapped. 

The data indicate that the flux increased by a significant
fraction compared to the uncertainties between 2003 and 2009. Such a
trend is indeed anticipated if the system is approaching
periastron passage c.f. WR\,140 \citep{Dougherty2005}.

 A synchrotron component that dominates over the thermal 
emission from the stellar winds of the two stars can explain the 
measured flux densities. We have modeled the data with a power-law 
spectrum with a low-frequency exponential cutoff.
We have computed two possible models. The first
model fits the data from January 2003 - December 2003, and the 2004
 data at the higher frequencies (Fig. 9: solid dots). It is
assumed that the flux does not vary significantly within this range of
time, given the separation of Aa to Ab in that period.  These data
are well-fit by a power-law, with a turnover around 1.4 GHz. However, the data from 2008 and 2009 are
modeled by a similar power-law spectrum but with a turnover at a slightly
higher  frequency.
An additional thermal component is required to fit the data at 20 GHz.
%
    \begin{figure}
   \centering
 \includegraphics[width=8cm,angle=0]{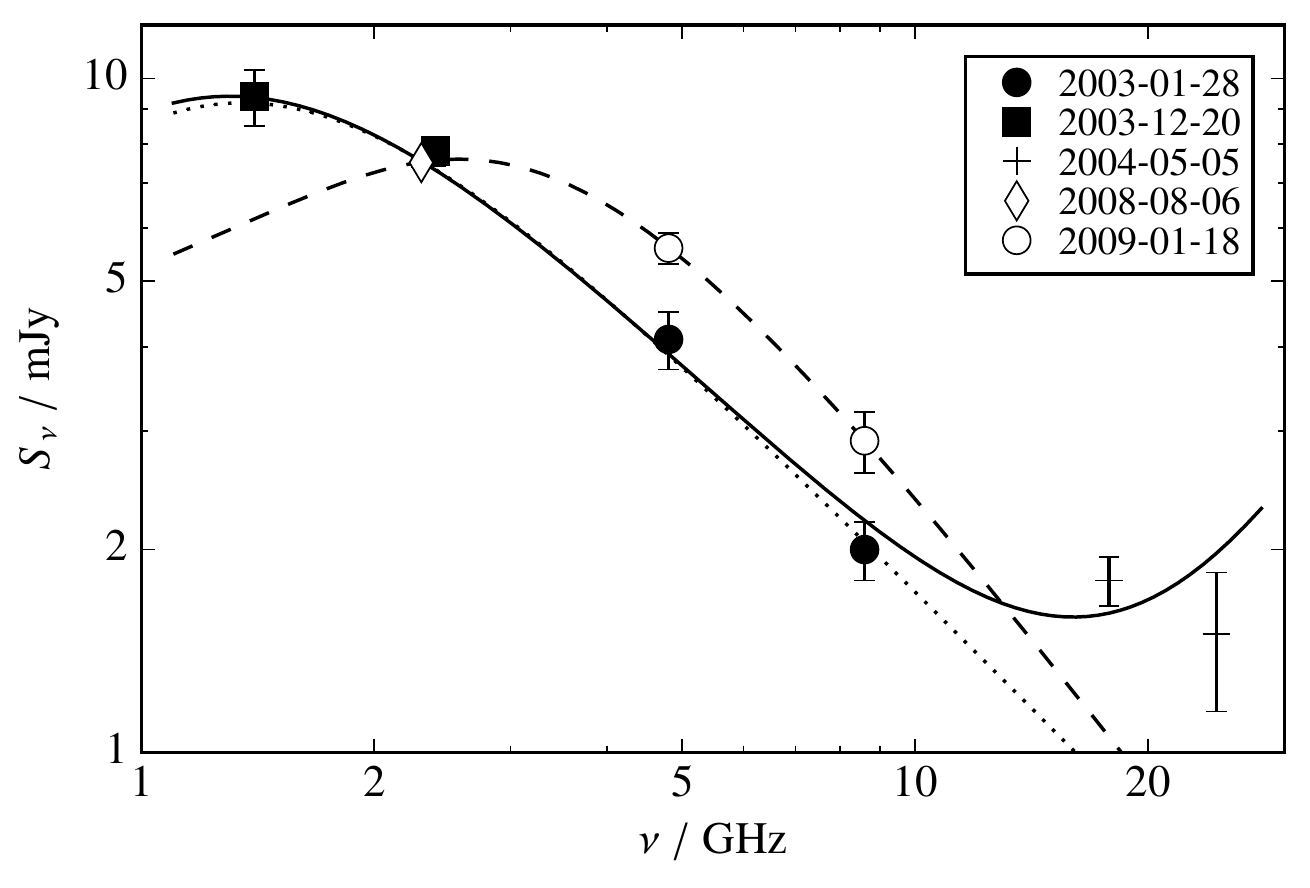}
      \caption{Spectra of HD~93129A obtained from ATCA data at different epochs (see
labels). The data from 2003 have been fitted with two different
models: a Free-Free Absorption (FFA, dotted line), and a FFA plus
a thermal emission also considering the 2004 data (solid line). The
2008--2009 data have been fitted with a FFA model (dashed line); see also Pittard \& Dougherty (2006).}
        \label{spectrum}
  \end{figure}
The spectral indices from the fits to the 2003-4 and 2008-9 data are, respectively,   
$-1.03 \pm 0.09$ and $ -1.21 \pm 0.03$. 

Up to now, the results from observations at high-energy spectral bands
remain inconclusive in detecting emission from a WCR.  In soft X-rays
(below 10 keV), the work by \citet{Cohen2011} concluded that the X-ray
spectrum is dominated by emission from the individual stellar winds in
the system. On the other hand, \citet{Gagne2011} reported a ratio
between the X-ray and bolometric luminosities 
($L_{\rm X} / L_{\rm bol} $) of 1.3\,$\times$\,10$^{-7}$.  
 This value suggests that only a small contribution of the X-rays come
from the colliding-wind region. It is not unexpected to see
colliding-wind binaries made of very early-type stars with no bright
X-ray emission in excess of the canonical $L_X/L_{\rm bol}$ value of
$\sim 10^{-7}$. For very early-type stars with strong and dense winds
such a the members of HD\,93129A, the individual X-ray emission should
be somewhat fainter than suggested by the canonical ratio (Owocki et
al.\,2013; De Becker 2013). A low emission measure of a WCR due to a
large stellar separation on top of such an underluminous wind emission
leads therefore to a not so high overall $L_X/L_{\rm bol}$ ratio.

HD\,93129A is rather close to the Fermi source 1FGL J1045.2$-$5942 (in
the 0.1-100 GeV energy range). This source is associated with
$\eta$\,Car, the only PACWB identified on the basis of $\gamma$-ray
emission. It seems very unlikely that HD\,93129A contributes to
the emission detected by Fermi, as clarified by \citet{Abdo2010}. At
even higher energies, no TeV source has been found at the position of
HD\,93129A in the TevCat catalogue (http://tevcat.uchicago. edu/),
even when considering newly announced sources and source
candidates. We conclude that there is no known $\gamma$-ray source
associated with HD\,93129A. Observations with the forthcoming
Cherenkov Telescope Array, with unprecedented sensitivity and source
location techniques, will surely  provide useful constraints to 
better understanding of the high energy phenomena in these objects.

\section{Summary and conclusions}\label{conclusions}

An LBA observation of the massive binary HD\,93129A detected an
extended and arguably curved radio emission component with a flux
density of 2.9 mJy at 2.3 GHz between the two components Aa and
Ab. Following a detailed analysis of recent high angular resolution
astrometry of the system, we provide compelling evidence that the
radio emission component is coincident with the expected position of a
wind interaction region between components Aa and Ab, and suggest a
wind-momentum ratio of $\sim0.5$.

Historic ATCA observations of the HD\,93129A system, from 1 to 25 GHz,
re-reduced in a uniform way,  show that the flux increased between
2003-4 and 2008-9, both epochs being well fit by a power-law spectrum with a 
steep spectral index. Similar increases in flux have been seen in other systems as they
approach periastron.  Thus, the results presented in this paper lend
significant additional support to the idea that wind-collision regions 
are the sites where relativistic particles are accelerated in 
particle-accelerating colliding-wind binaries.

Very Long Baseline Interferormeter (VLBI) observations of particle-accelerating
colliding-wind binaries help to quantify specifically the properties
of the non-thermal radio emission. 
Such measurements are crucial for models aimed at reproducing the
particle acceleration and non-thermal physics at work in these
objects.  VLBI observations of the HD 93129A system across a wide
frequency range will allow more accurate properties of the spectrum to
be derived, and hence the properties of the relativistic electron
populations involved in the synchrotron emission process, and
potentially reveal associations between flux variations and changes in
the properties of the wind-collision region.

In parallel, new ATCA observations, preferentially simultaneous across
all the observing frequencies are necessary to obtain accurate 
total fluxes. Repeated observations in the leading up to the putative periastron passage
will reveal if the flux increases as observed in other PACWB systems at this
phase of the orbit. 

\begin{acknowledgements}
The authors thank an anonymous referee for detailed comments, C. Phillips and A. Tzioumis for their work in a
preliminary stage of the study, A. López-Sánchez for providing some of the ATCA observations, 
Rodolfo Barb\'a, Roberto Gamen, Ed
Fomalont, Jamie Stevens, and very
especially G. Vila and N. Casco for help with the orbits.
P.B. is
supported by the ANPCyT PICT- 2012/00878 and UNLP G11/115 project.
B.M. acknowledge support by the Spanish Ministerio de Econom\'ia y
Competitividad (MINECO) under grants AYA2013-47447-C3-1-P and
BES-2011-049886. This study was partially supported by NASA through
grant GO-10898 from the Space Telescope Science Institute, which is
operated by AURA, Inc., under NASA contract NAS 5-26555. The research
has made use of the Washington Double Star Catalog maintained at the
U.S. Naval Observatory.
\end{acknowledgements}

\end{document}